# Quantitative EUV ptychography reveals nanoscale morphological responses of bacteria under physiological and antibiotic stress

*Chang Liu[1,2,3,*], Leona Licht[1,2,3], Christina Wichmann[4,5,6], Wilhelm Eschen[1,2,3,§], Soo Hoon Chew[1,7], Felix Hildebrandt[4,5], Daniel S. Penagos Molina[1,2,3], Christian Eggeling[4,5,6], Jens Limpert[1,2,3,7] and Jan Rothhardt[1,2,3,7]*

** Corresponding author, E-mail: liu.chang@uni-jena.de*

*1. Institute of Applied Physics, Abbe Center of Photonics, Friedrich-Schiller-University Jena, Albert-Einstein-Straße 15, 07745 Jena, Germany*

*2. Helmholtz-Institute Jena, Fröbelstieg 3, 07743 Jena, Germany*

*3. GSI Helmholtzzentrum für Schwerionenforschung, Planckstraße 1, 64291 Darmstadt, Germany*

*4. Institute of Applied Optics and Biophysics, Friedrich Schiller University Jena, Philosophenweg 7, 07743 Jena, Germany*

*5. Leibniz Institute of Photonic Technology (Leibniz-IPHT), Albert-Einstein-Straße 9, 07745 Jena, Germany; Member of the Leibniz Center for Photnics in Infection Research (LPI), Jena, Germany*

*6. Cluster of Excellence Balance of the Microverse, Friedrich Schiller University Jena, Jena, Germany*

*7. Fraunhofer Institute for Applied Optics and Precision Engineering, Albert-Einstein-Str. 7, 07745 Jena, Germany*

*§ current affiliations: JILA, 440 UCB, Boulder, CO 80309, USA*

**Abstract:**

Table-top extreme ultraviolet (EUV) ptychography enables nanoscale, label-free, and quantitative imaging with intrinsic elemental sensitivity, offering a unique modality for subcellular profiling of bacterial morphology and composition. In this work, we apply a state-of-the-art EUV ptychographic microscope to systematically investigate the structural and compositional features of two model prokaryotic bacteria, *Escherichia coli* and *Bacillus subtilis.* With quantitative amplitude and phase reconstructions at 44 nm resolution on a tabletop, we visualize subtle phenomena during bacterial sporulation based on distinct morphological signatures. Notably, we examine the single-cell response of *B. subtilis* to the antibiotic monazomycin, uncovering

ultrastructural disruption and compositional alterations. To further characterize the phenotypic variations, we perform multivariate statistical analysis on extracted morphological features, revealing diversity and clustering patterns associated with defined biological states. This work establishes EUV ptychography as a powerful, element-sensitive imaging platform for label-free bacterial imaging, showcasing its promise for biomedical and antimicrobial research.

**Introduction:**

Bacterial infections pose a significant threat to human health, driving the demand for precise diagnostic tools and in-depth characterization of bacterial structures, physiological states, and responses to antimicrobial agents[1,2]. While mature molecular techniques such as polymerase chain reaction (PCR)[3], mass spectroscopy[4], and Raman-based spectroscopy[5,6] provide valuable insights into bacterial identity, they lack spatial resolution and often require amplification or labeling[7]. Advanced microscopy methods, including fluorescence microscopy with super-resolution[8-10], atomic force microscopy (AFM)[11], magnetic resonance imaging (MRI)[12], soft X-ray microscopy[13-15], and cryo-electron microscopy[16-18], have provided valuable insights into bacterial ultrastructure. However, they each come with trade-offs: labelling procedures in fluorescence microscopy are often complex, time-consuming, and may introduce morphological artifacts or alter physiological states[19]. Besides, AFM and EM provide high-resolution imaging but suffer from limited depth information[20,21], while methods using X-rays are particularly prone to radiation-induced damage[22], complex sample preparation, or lack of compositional sensitivity[23]. These trade-offs highlight a pressing need for a compact, label-free, non-destructive imaging technique with excellent material contrast, capable of resolving subcellular bacterial morphological features.

Since the first successful reconstruction of a hard X-ray diffraction pattern taken from an intact *Escherichia coli* cell at the Spring-8 synchrotron facility[24], coherent diffractive imaging (CDI)[25] with X-ray (100 eV - 100 keV) light has opened new avenues for nanoscale biological imaging[23]. Unlike conventional microscopy, CDI, and in particular its scanning variant, ptychography[26], computationally reconstructs both the amplitude and phase of the transmitted or reflected wavefield from a series of diffraction patterns. Ptychography has attracted considerable attention in biological studies across X-ray[27-31] and electron[32-34] beams, providing quantitative, sub-30 nm images of cells and extended organelles under cryogenic conditions. In addition to short-wavelength implementations, ptychography has also been demonstrated in the visible light regime, enabling label-free live-cell biomedical imaging with enhanced phase contrast[35,36].

At photon energies in the extreme ultraviolet (EUV, 10 - 100 eV) range, ptychographic imaging has been widely applied in nanotechnology[37,38], wavefront sensing[39], and reflectrometry[40], but its biological applications remain limited due to strong EUV absorption in hydrated samples. Nonetheless, EUV ptychography offers several unique advantages for dried biological specimens: the short wavelengths allow for nanoscale spatial resolution[41], while numerous atomic resonances, especially of the biologically relevant elements carbon, nitrogen, oxygen, and phosphorus, provide strong, intrinsic elemental contrast[42] in dehydrated cells without the need for labeling or staining. Notably, the capability of ptychography to simultaneously retrieve amplitude and phase enables dose-efficient imaging in the EUV regime compared with traditional X-ray microscopy[43], achieving nanoscale resolution at radiation levels well below the well-known Henderson limit[22].

Recent advances in high-flux laboratory EUV sources[44] based on high harmonic generation (HHG)[45] have facilitated compact EUV ptychographic microscopes[46], which have successfully imaged dehydrated mouse neurons[47], fungi, and bacteria[48]. However, these demonstrations were constrained by moderate spatial resolution (~ 80 nm)[47], low throughput (< 1 Mpixel/h)[47,48], and reconstruction artifacts[48] due to model mismatch in forward propagation algorithms, restricting their sensitivity and quantitative accuracy for biological analysis.

To enable reliable quantitative imaging, we employed an upgraded EUV ptychographic microscope operating at 92 eV[36], featuring optimized illumination optics, a low-noise sCMOS detector[49], and a newly developed purity-based autofocusing algorithm[50]. This system achieves a 44 nm resolution on dried biological samples with imaging speed improved by a factor of four compared to the previous EUV imaging studies on biological samples[47,48]. The utilized ptychographic microscope is described in *Methods*.

Building on this capability, we present a systematic EUV ptychographic study of two model bacterial species, *Escherichia coli* and *Bacillus subtilis*, captured as unstained, air-dried samples without chemical fixation. Next, we investigate morphological and compositional changes during *B. subtilis* sporulation[51], as well as the cellular response to monazomycin, a membrane-targeting macrolide antibiotic[52]. The reconstructed phase and amplitude are further processed to extract a scattering quotient[28] metric, enabling label-free elemental mapping of bacterial ultrastructure.

Notably, we apply feature-based multivariate statistical analysis to EUV image datasets for the first time, enabling phenotypic profiling of individual bacterial cells treated with monazomycin. This approach reveals distinct clustering patterns corresponding to defined physiological states, demonstrating the quantitative capability of EUV ptychography.

Altogether, our results establish EUV ptychography as a promising platform for high-resolution, label-free, quantitative morphological and compositional imaging in microbiological and antimicrobial research.

## Results

### *Morphological imaging of two representative bacterial models*

To initiate our investigation, we selected two widely-studied bacterial models: *Escherichia coli* (*E.coli*, Gram-negative) and *Bacillus subtilis* (*B. subtilis*, Gram-positive), which, while comparable in size and rod-like shape (typically 2–5 μm in length, 0.2-1 μm in diameter[53]), exhibit pronounced differences in their cell wall architecture. *E. coli* features a thin peptidoglycan layer (2 - 7 nm) enclosed by an outer membrane enriched with lipopolysaccharides (LPS) and lipoproteins, whereas *B. subtilis* is encapsulated by a substantially thicker peptidoglycan wall (20 - 80 nm) enriched with lipoteichoic acid (LTA) and teichoic acids[54].

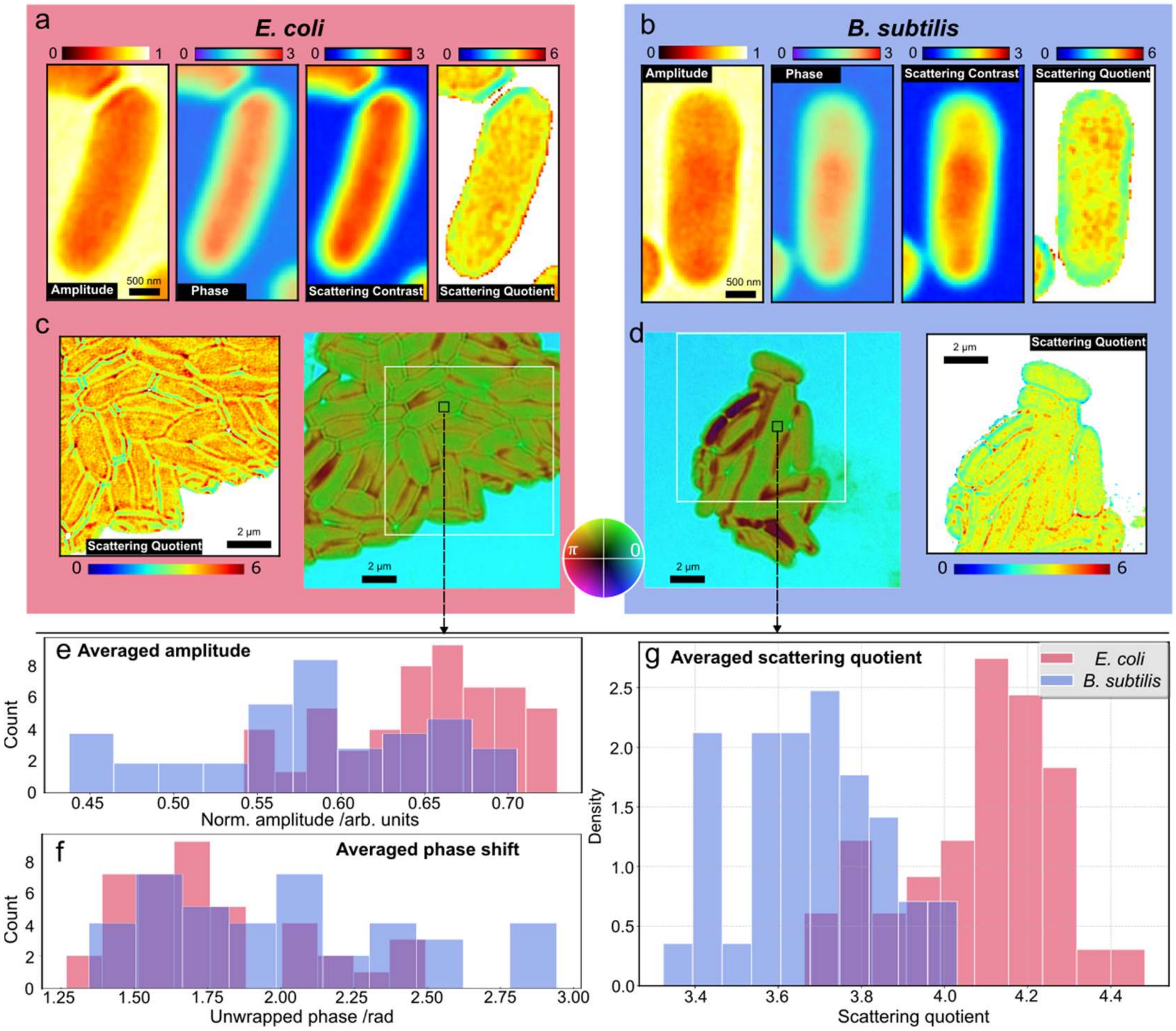

**Figure 1 EUV ptychographic imaging of *E. coli* and *B. subtilis* reveals morphological and compositional contrast at the single-cell and population level. a, b.** Complex-valued EUV reconstructions of single *E. coli* (**a**, red box) and *B. subtilis* (**b**, blue box) cells, showing amplitude, unwrapped phase, scattering contrast, and scattering quotient, respectively. **c, d.** Wide-field scattering quotient maps of bacteria clusters of multiple *E. coli* (**a**) and *B. subtilis* (**b**) cells within selected regions of interest (ROIs) in complex-valued transmission (white boxes). Brightness and hue encode modulus and phase, respectively. Black boxes (each 400 nm in size) indicate the ROIs used for quantitative analysis shown in (**e-g**), with 10 ROIs selected per image. **e-g** Histograms of mean values extracted from individual ROIs (n = 40 per species) for (**e**) normalized amplitude, (**f**) unwrapped phase shift, and (**g**) scattering quotient, showing quantitative distributions and statistically distinct biophysical properties between the two species. Color bars are normalized per channel. Scale bars: 500 nm in (**a, b**), 2 μm in (**c, d**).

EUV ptychographic measurements were performed on two independently prepared samples per bacterial species at sub-50 nm half-pitch resolution, covering multiple regions of interest (ROIs) to ensure robustness against biological variability and imaging conditions. The complete workflow and all reconstructed fields of view (FOV) are summarized in Supplementary Figure S2. Sample preparation details are provided in the *Methods*.

Figure 1**a** and **b** show representative reconstructions of individual *E. coli* and *B. subtilis* cells, respectively. These two species differ fundamentally in their envelope architecture (Supplementary Figure S1). The amplitude ($A$) and unwrapped phase ($\varphi$) images reflect the cell's absorption and refraction properties. From these, scattering contrast[55] ($f_c = \sqrt{\ln^2 |A| + \varphi^2}$) and scattering quotient[28] ($f_q = \varphi / \ln A$) maps are extracted to emphasize scattering strength and material contrast, respectively. As reported previously[28,36,48], these integrated metrics help reveal ultrastructural features less apparent in amplitude or phase alone.

The $f_c$ map shows variations in envelope morphology, with the *E. coli* cell appearing more layered and sharply defined, while *B. subtilis* cell displays a more diffuse internal profile. The $f_q$ map, which uncovers the projected elemental composition averaged along the optical axis, further emphasizes the distinctions. For instance, *E. coli* cell exhibits a high-contrast, single-pixel-wide shell at the outer boundary ($f_q > 5$), which is nearly absent in *B. subtilis*. This peripheral enhancement may be consistent with the presence of a lipid- and LPS-rich outer membrane ($f_q \approx$ 6.0 for lipids at 92 eV, see Supplementary Table 1), commonly reported in Gram-negative bacteria[56]. Moreover, the *E. coli* displays generally higher and more uniform $f_q$ values within the cell interior, whereas *B. subtilis* presents a localized $f_c$ hotspot with relatively lower $f_q$ values, which could be attributed to carbohydrate-rich structures of the thick peptidoglycan layer in Gram-positive bacteria ($f_q \approx 3.5$ for carbohydrates at 92 eV, see Supplementary Table 1).

To account for cell-to-cell variability, a population-level comparison was conducted. Figure 1**c** and **d** present complex transmission images of representative clusters, with hue encoding the phase shift and brightness representing amplitude attenuation. Corresponding $f_q$ maps of zoom-in FOVs imply the compositional difference between the two species. For statistical analysis, a total of 80 ROIs (40 per species) were selected from 8 independent measurements, focusing on cellular regions (black boxes). The distributions of mean amplitude, phase, and $f_q$ values across these ROIs are shown in Figure 1**e-g**. Notably, the $f_q$ histogram reveals a partially separated distribution, suggesting compositional trends consistent with the expected biochemical components in the cell wall of each species. A more detailed interpretation, along with reference $f_q$ values for key biochemical components, is provided in the Supplementary.

Taken together, the EUV ptychographic reconstructions reveal reproducible morphological and compositional contrasts between *E. coli* and *B. subtilis*, with population-level trends consistent with expected envelope biochemistry. While not intended as a species-classification assay, these label-free measurements provide subcellular sensitivity to envelope architecture and composition, complementing established methods such as Gram staining, which differentiates bacteria by cell-wall composition, particularly the thickness of the peptidoglycan layer[57], but lacks subcellular resolving power, thereby underscoring the complementary value of EUV ptychography.

***Visualization of B. subtilis sporulation***

Endospore formation is a well-known survival strategy adopted by certain bacterial species to endure harsh environmental conditions[51]. These dormant structures protect the bacterial genome during periods of stress, such as desiccation, freezing, high temperature, toxic chemicals, or intense radiation exposure, including UV and γ radiation[58]. Specifically, *B. subtilis* serves as a widely studied model organism for spore formation, relevant to both food safety and antimicrobial research[59]. While extensive genetic and biochemical investigations have been conducted, structural visualization of sporulation has primarily relied on electron microscopy[60] and X-ray[61] illumination, which often require elaborate sample preparation and a lack of phase contrast.

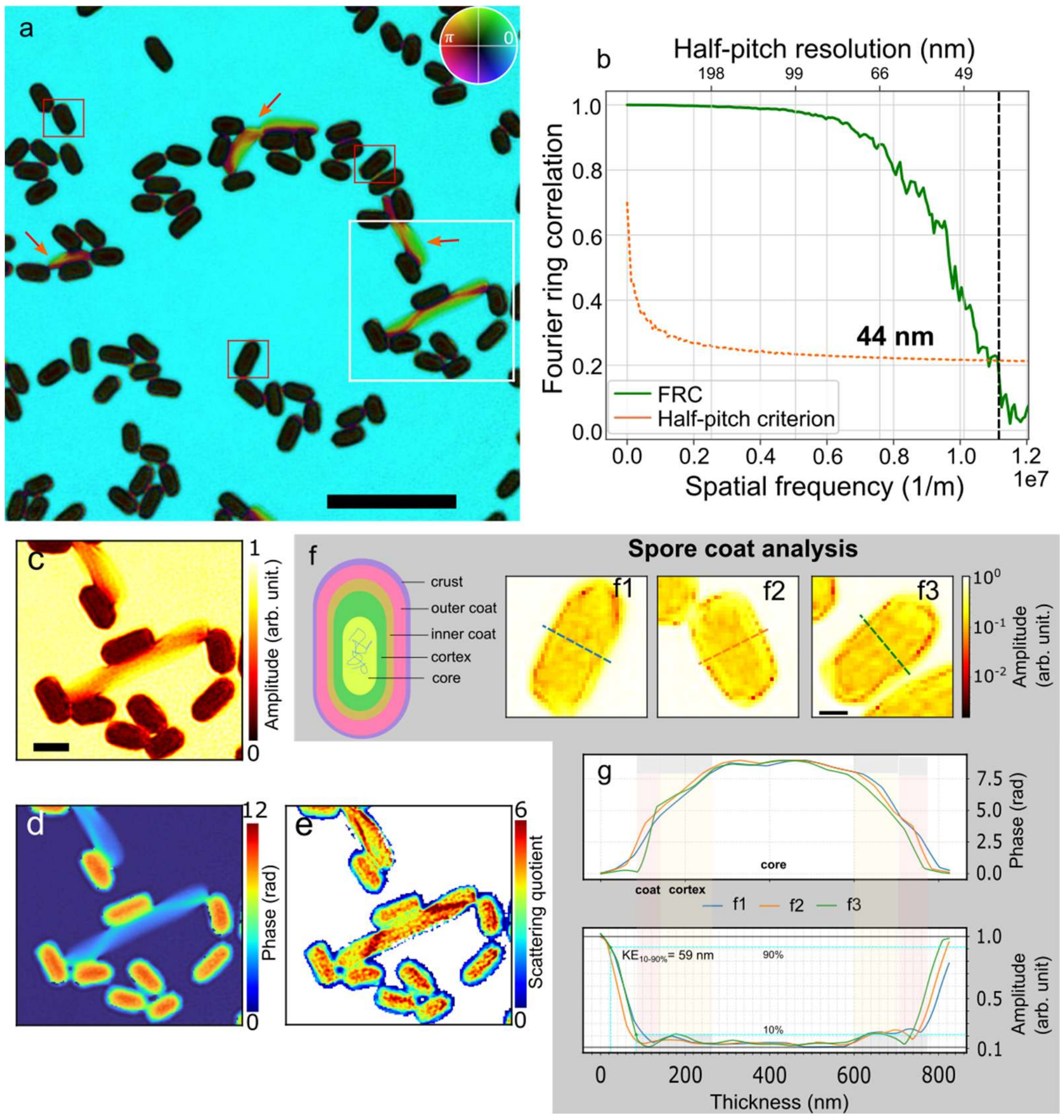

**Figure 2 EUV ptychography characterizing *B. subtilis* sporulation. a.** Complex-valued transmission image of *B. subtilis* under sporulation. Brightness and hue encode modulus and phase, respectively. Scale bar: 5 µm. **b.** Fourier ring correlation analysis of the EUV image **a**, showing a spatial resolution of 44 nm. **c**, **d**, and **e** represent the normalized, vacuum-referenced amplitude, unwrapped phase, and scattering quotient of the ROI indicated by the white box in **a**, respectively. Scale bar: 1 µm. **f**. Schematic model of *B. subtilis* spore coat structure. **f**1 - **f**3 Logarithmic amplitude maps of three individual spores (red box in **a**), revealing concentric amplitude modulations corresponding to spore coat layers. Scale bar: 300 nm. **g.** Lineout profiles from spores in **f**1-**f**3 showing the separation between coat and cortex. A knife-edge test yields a 10-90% amplitude transition over ≈59 nm.

In this study, *B. subtilis* samples are treated with $MnSO_4$ to create conditions favourable for sporulation[62] and imaged using EUV ptychography. The long-term culture results in nutrient depletion and a heterogeneous mixture of bacterial cells and spores, which can be distinguished by morphological and scattering variations. As shown in Figure 2**a**, the reconstructed complex image reveals rod-shaped cells and ellipsoidal structures corresponding to individual spores. A resolution analysis using two independent measurements and a Fourier ring correlation (FRC)[63] indicates a diffraction-limited half-pitch resolution of 44 nm (Figure 2**b**).

Next, ROIs containing cells and spores are selected for detailed analysis. A rod-shaped cell (white box in Figure 2**a**) is seen in proximity to four adjacent ellipsoidal spores. The amplitude, phase, and scattering quotient ($f_q$) maps of this region are shown in Figure 2**c-e**. While the amplitude and phase suggest the spore is denser than the cell, the scattering quotient remains relatively uniform, indicating similar average composition. By contrast, the filament structure connecting multiple spores displays a sharp peak in both phase and scattering quotient. This structure may correspond to an extracellular matrix component or hydrophobic residue[64], although its identity remains speculative. Additionally, other cells (red arrowheads in Figure 2**a**) appear to exhibit morphological abnormalities, such as localized swelling, division, and irregular cellular contours.

To further verify the resolving power of our microscope, we selected three individual spores for coat analysis. The apparent concentric layering observed in the logarithmic amplitude maps (Figure 2**f1**-**f3**) reveals nanoscale coat envelope structures. A 10-90% knife-edge analysis of the representative profile **f3** (green curve in Figure 2**g**) yields a width of 59 nm, consistent with the FRC-derived 44 nm resolution when accounting for the intrinsic gradient of biological coat boundaries. Line profiles across each spore consistently exhibit a narrow high-absorption ridge at the periphery (red shading), followed by a lower-absorption band (yellow shading) and finally a homogeneous core region. This pattern agrees with the established *B. subtilis* spore coat assembly[65]. The outer high-absorption ridge corresponds to the protein-rich coat, approximately 80 nm[66], which exhibits elevated EUV absorption. Beneath the coat lies the cortex, a protective peptidoglycan-rich compartment that appears as an intermediate band of ~100 nm width[67]. The thickness of the outermost polysaccharide crust layer[68] lies below our resolving power and is not expected to be individually resolved.

These observations show that EUV ptychography enables the visualization of representative morphological diversity potentially associated with sporulation, providing sub-50 nm, label-free contrast that offers new insights into the subcellular characteristics of *B. subtilis* spores. Future efforts will focus on time-resolved EUV imaging on *B. subtilis* across successive stages of sporulation, aiming to capture morphological transitions and establish reliable stage assignments incorporating well-defined sporulation mutants (e.g., cotE[69]), potentially supported by fluorescence-based references in future correlative stages.

***Visualization of bacterial responses to antibiotic monazomycin***

Antibiotics are potent medications that either inhibit bacterial reproduction or kill bacteria outright. The latter antibiotics often target essential processes, such as bacterial cell wall synthesis or interference with cellular contents like DNA, proteins, or metabolic pathways[70].

Membrane-targeting antibiotics such as monazomycin are widely used against Gram-positive bacteria. As a polyether macrolide, monazomycin intercalates into lipid bilayers and forms hydrophilic pores, leading to membrane depolarization and uncontrolled ion leakage[71]. Unlike β-lactams that target peptidoglycan synthesis, the biochemical mechanism of monazomycin has not been extensively studied and remains virtually unexplored. To our knowledge, no microscopy-based investigations have been reported for monazomycin-treated Gram-positive bacteria. While various imaging studies employing fluorescence[72,73], AFM[74], and mostly EM[75-78] have revealed morphological responses to other antibiotics and bacterial species, these approaches often lack sufficient spatial resolution or elemental specificity and cannot be directly extrapolated to monazomycin due to its distinct membrane-targeting mechanism.

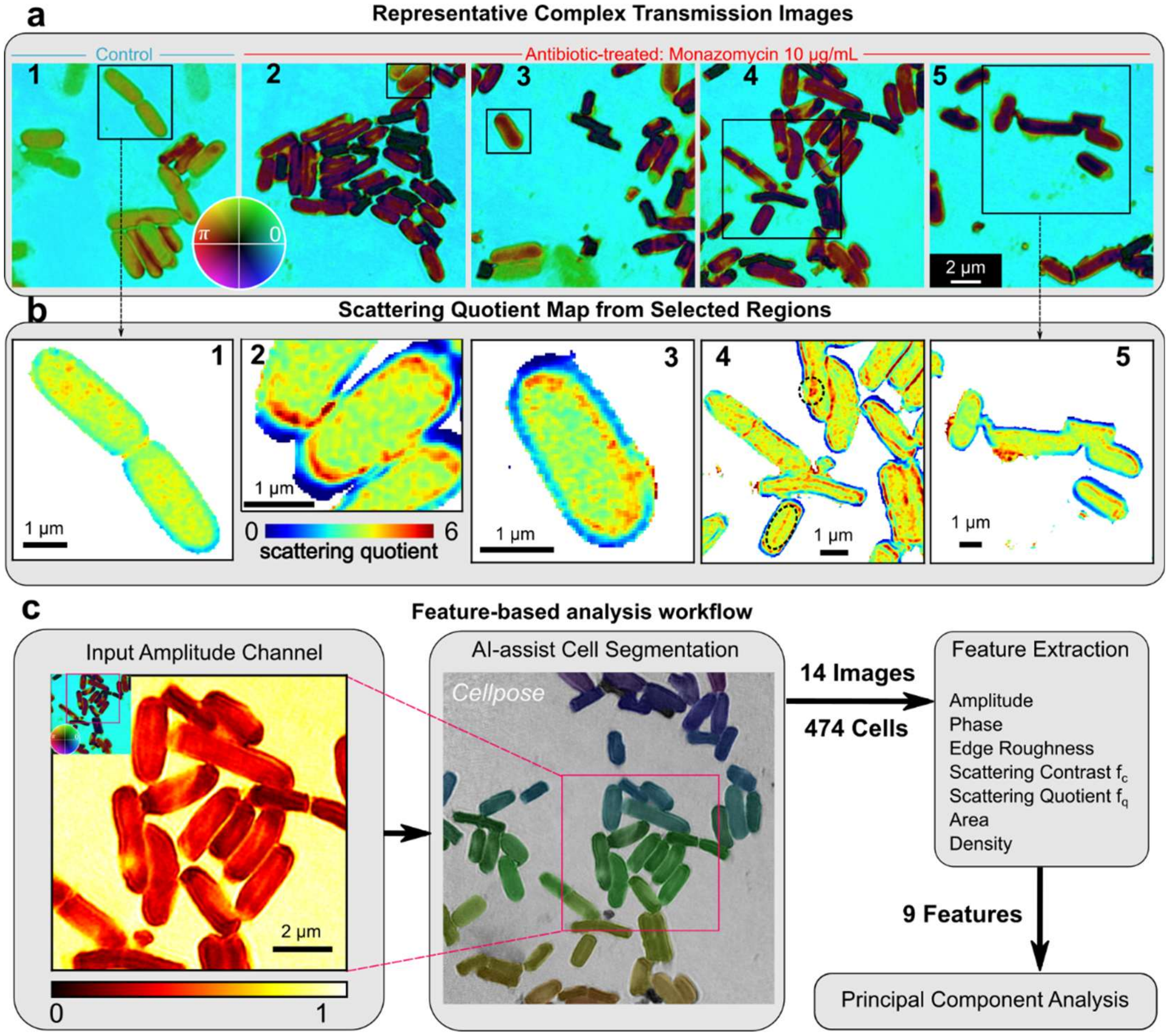

**Figure 3 EUV ptychographic imaging and postprocessing of *B. subtilis* under antibiotic treatment. a.** Representative complex-valued transmission images of control (1) and 10 µg\mL monazomycin-treated cells (2-5), with insets marking cells for downstream analysis. Brightness and hue encode modulus and phase, respectively**. b**. Zoom-in scattering quotient ($\boldsymbol{f_q}$) maps highlight material diversity and change upon treatment. **c**. Cell segmentation and feature analysis workflow. From the input amplitude channel, cells from 14 independent EUV images are segmented using the CellPose-SAM algorithm, and masks are generated. Multiple morphological and scattering-based features are then calculated from each cell for quantitative analysis via principal component analysis (PCA).

Here, we employ EUV ptychography to visualize the morphological diversity of *B. subtilis* following monazomycin treatment. By capturing a wide spectrum of structural phenotypes, including membrane fragmentation, condensation, and swelling, we reveal a highly heterogeneous cellular response at the single-cell level.

Identical cultures of *B. subtilis* were cultivated under standardized conditions, with one group treated with 10 µg/mL monazomycin and the other serving as an untreated control. Both groups were subsequently prepared on SiN membranes and imaged with EUV ptychography under identical acquisition parameters, enabling a direct comparison of morphological and structural responses to monazomycin. Figure 3**a** displays complex transmission reconstructions from representative FOVs for both conditions. Compared to the untreated controls, treated cells exhibit notable morphological alterations, including increased scattering density, irregular shape deformation (**a**3 - **a**5), and reduced internal uniformity. In contrast, control cells largely retain their rod-like morphology and structural integrity. These structural changes are further supported by scanning electron microscopy (SEM) images provided in the Supplementary, which visually confirm disruption with lower contrast. In Figure 3**b**, $f_q$ maps of the selected ROIs marked in Figure 3**a** further reveal pronounced features in the element-contrast-enhanced ultrastructure. In particular, a distinct shell-like layer with $f_q > 5$ is frequently observed near the periphery of treated cells (**b**2 - **b**5). This layer, absent in controls (**b**1), likely reflects the formation of a disordered or detached membrane bilayer due to the monazomycin.

Next, to enable large-scale quantitative analysis of the cellular morphological and compositional response, a pipeline of single-cell feature-based multivariate analysis based on numerous EUV reconstructions was implemented, as illustrated in Figure 3**c**. The reconstructed amplitude channel was used as input for the deep learning-based segmentation algorithm Cellpose-SAM[79]. The fully automated method requires no model retraining or parameter adjustment and accurately segments typical rod-like bacterial cells. A total of 474 cells were segmented from 14 representative reconstructions in both controlled (4) and treated groups (10). For each segmented cell, a comprehensive set of nine features was extracted to capture structural morphology, edge roughness, and compositional characteristics. These include the statistical descriptors such as mean, standard deviation (std), and relative standard deviation (RSD) from both amplitude and phase channels, edge roughness calculated from amplitude and phase, scattering metrics ($f_c$ and $f_q$), projected cell area, and estimated density. The scattering-related features offer material-specific information while roughness and density serve as morphological indicators of cellular integrity and uniformity. The resulting feature vectors were subjected to three-dimensional principal component analysis (PCA) for dimensionality reduction. While several extracted features are derived from the reconstructed complex transmission function and thus exhibit strong intercorrelations, PCA effectively captures the dominant modes of variation while minimizing redundancy, as illustrated in Figure 4**a**. The first two principal components account for 52.77% (PC1) and 20.25% (PC2) of the total variance, which reveal distinct distributions between treated and control groups along these two axes. The relative contributions (loadings) of each feature to PC1 and PC2 are shown in Figure 4**b**. $f_c$ density and roughness-related features contribute most strongly to PC1, reflecting that structural complexity and heterogeneity vary the most within those mutants. In contrast, the orthogonal axes PC2 are dominated by the projected area and RSD of $f_c$.

In addition, a statistical comparison of single-cell features between control and treated groups is presented in Figure 4**c** as standardized violin plots, ranked by statistical significance. Notably, $f_c$ *density* and *phase mean* exhibit the most pronounced differences, indicating a consistent deviation in intracellular material content upon treatment. These changes likely reflect membrane disruption and cell condensation. Similarly, *amplitude roughness* is significantly increased in treated cells, indicative of irregular scattering surfaces, possibly due to membrane disorganization. It is also worth noting that *RSD of* $f_c$, a measure of spatial variation in scattering properties, is significantly increased as well, supporting the hypothesis of disrupted internal uniformity under treatment. Lastly, the *projected area* exhibits a broader distribution in the treated group, suggesting that monazomycin may induce alteration of the cell size.

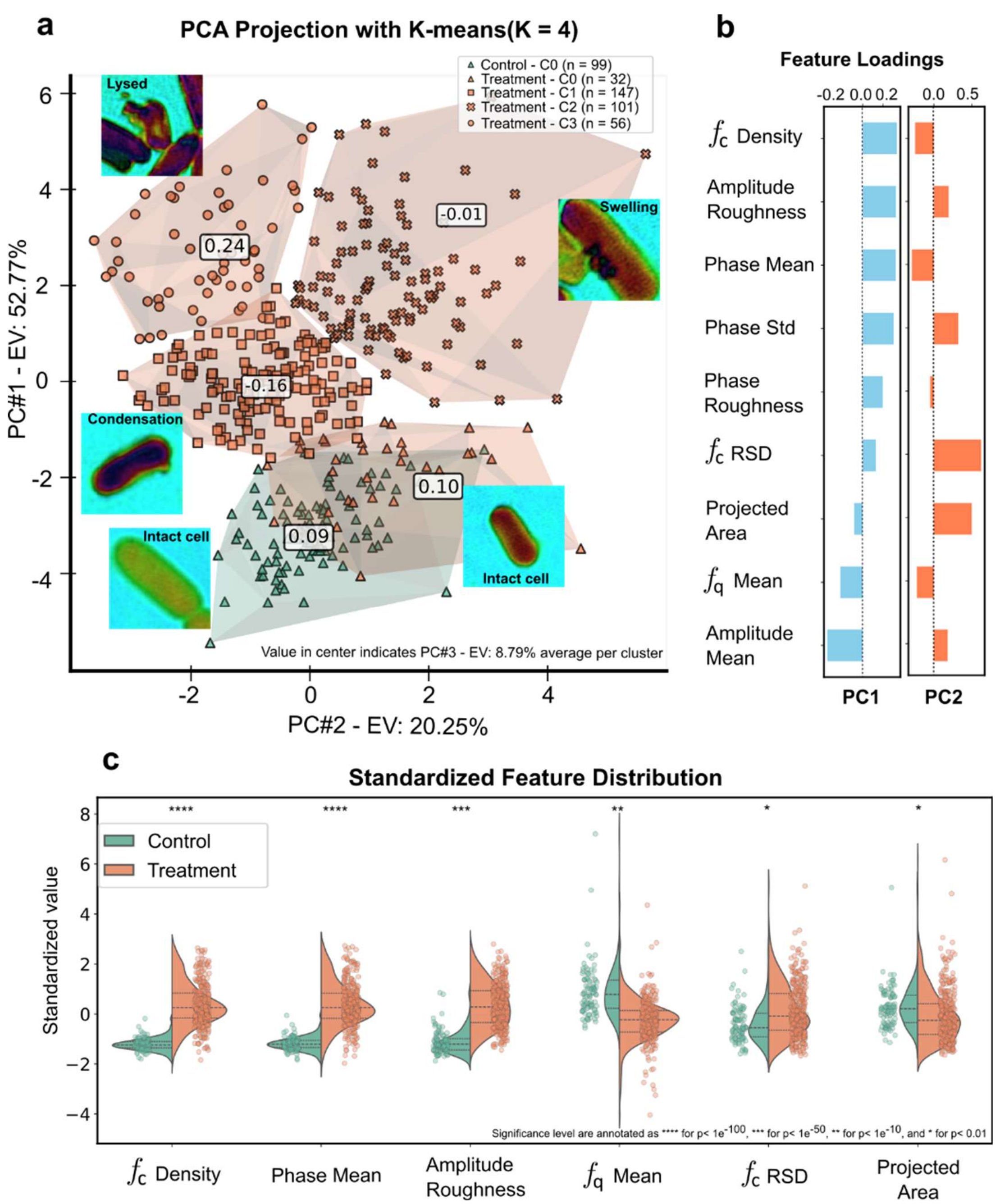

**Figure 4 Quantitative phenotyping of *B. subtilis.* cells based on EUV-derived single-cell features. a.** Three-dimensional PCA of 474 segmented cells using nine extracted features, projected in PC1-PC2 space. Four major clusters are identified through unsupervised K-means clustering (K = 4), enclosed by convex hulls. The number of clusters was determined by silhouette analysis[81]. Each point represents an individual cell, colored by treatment condition. Cluster centers are labeled with their mean PC3 value. Representative cells for each cluster centroid are shown as insets. **b** PCA loadings indicating the contribution of each parameter to PC1 and PC2, sorted by PC1 importance. Std: standard deviation. RSD: relative standard deviation. **c** Violin plots comparing features distribution between the control and antibiotic-treated groups. Features are sorted by statistical significance (***p***-value) between control and treatment groups using a two-sided Welch's t-test (unequal variance t-test).

To further categorize the phenotypic profiles of the 474 cells, we performed a similarity-based unsupervised clustering using the K-means[80] algorithm. The medium cluster silhouette score[81] (0.35 for K=4) indicates moderate separation, consistent with partially overlapping phenotypic clusters along a continuum of morphological states rather than sharply distinct categories. As illustrated in Figure 4**a**, the clusters exhibit the separation in the PCA space (projected onto PC1 and PC2), each enclosed by a convex hull.

Cluster 0 predominantly comprises untreated control cells (n = 99) along with a small number of minimally affected treated cells (n = 32). It forms a compact region characterized by low phenotypic variation. Representative control cells (Figure 4**a**, bottom left; Figure 5**a**, top row) exhibit intact rod-shaped morphology with uniform density and smooth boundaries. Treated cells within this cluster show slight swelling but retain optical features similar to control, suggesting an early stage of antibiotic exposure. Clusters 1 - 3 consist exclusively of treated cells and show more dispersed distributions in PCA space. Cluster 1 (n = 147) is marked by increased amplitude roughness and phase heterogeneity, with representative phenotypes showing signs of peptidoglycan thickening and separation of cell envelope layers[75]. In contrast, Cluster 2 (n = 101) shifts further along PC1 and PC2, and contains phenotypes with enlarged projected areas and diminished phase contrast, indicative of swelling, membrane bulging, and blebbing[76], morphological features previously reported primarily in Gram-negative bacteria via TEM and SEM[77,78]. Cluster 3 (n = 56) occupies the extreme positive end of PC1, representing phenotypes with pronounced amplitude roughness and a complete loss of phase integrity. These features are consistent with severe envelope rupture and cytoplasmic leakage, representing a terminal lytic state. Notably, Clusters 1 - 3 are composed entirely of monazomycin-treated cells, revealing a continuous phenotypic progression from intact to highly disrupted states. In comparison to representative images of the control group in Figure 5**a**, images in Figure 5**b** show increasing morphological and biophysical deterioration, such as bulging formation, blebbing, condensation, altered cell size, filamentation, and membrane lysis, in agreement with previously established antibiotic-induced phenotypes[83]. Notably, we also observed cells with almost completely depleted internal density yet preserved envelope integrity (Figure 5**b**, 'ghost cell' is used to describe those lysed bacteria[84]), indicative of complete cytoplasmic loss through polar leakage without gross membrane rupture. The preservation of cell shape may originate from mechanical constraints within dense cell clusters, where adjacent cells and the extracellular matrix prevent envelope collapse despite cytoplasmic depletion.

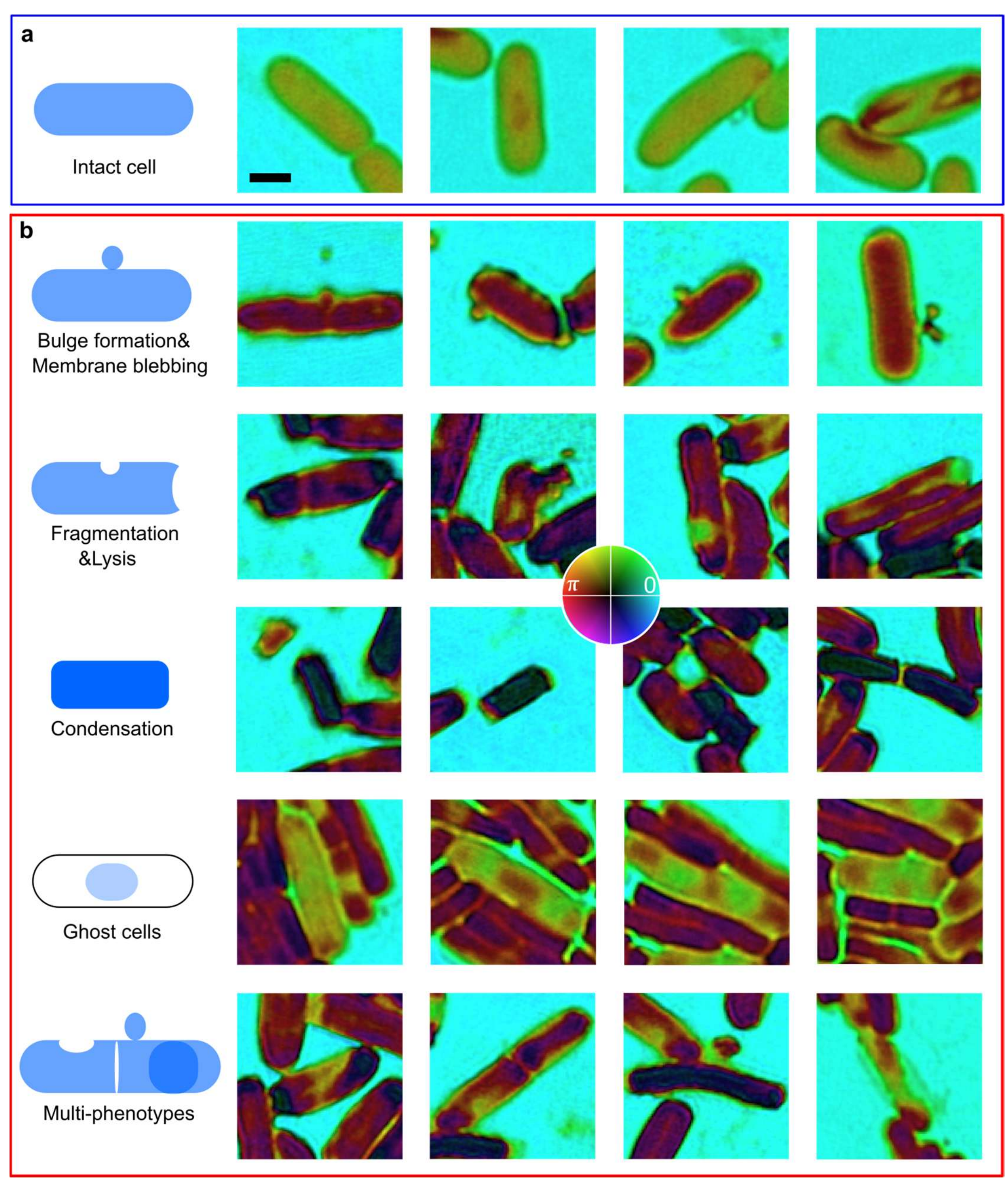

**Figure 5 EUV phenotypic visualization of representative *B. subtilis* cells under monazomycin treatment. a.** Negative control cells. **b** Antibiotic-treated cells displaying a wide range of damage phenotypes. Scale bar: 1 μm.

Overall, our results revealed that monazomycin elicits not a binary but a graded response in bacterial cells, ranging from early-stage membrane detachment to full cytolysis, consistent with previously reported phenotypic continua induced by various kinds of antibiotics.

***Discussion and outlook***

In this work, we demonstrate that EUV ptychography enables quantitative, label-free imaging of bacterial cells across multiple biological contexts at sub-50 nm resolution, including different bacterial species, developmental states, and antibiotic-induced disruption. The achieved resolution is estimated via FRC and independently cross-validated by high-resolution SEM comparison of the same bacterial ultrastructure (Supplementary Note 2).

As a baseline study, the comparative imaging of Gram-positive (*B. subtilis*) and Gram-negative (*E. coli*) species reveals pronounced differences in scattering quotient maps, attributable to their distinct cell wall compositions, including peptidoglycan thickness and lipid content, and illustrates the material specificity of EUV ptychographic imaging. Furthermore, the visualization of *B. subtilis* sporulation reveals subcellular morphological features and variations at 44 nm resolution, including the characteristic high-absorption peripheral ridge across the spores, and a filament along the cell. While the biochemical origin of these features remains to be fully established, their consistent appearance across multiple spores highlights the sensitivity of EUV ptychography to subtle subcellular changes. This capability opens the door to future quantitative, label-free investigations of sporulation dynamics and other bacterial developmental processes.

Most importantly, the study of the response of *B. subtilis* to monazomycin treatment reveals a continuum of phenotypic states. This gradient is supported by feature-based clustering and visual phenotyping, exposing a spectrum of morphological deterioration from early membrane detachment to cytoplasmic leakage and lysis. The ability to resolve a continuum of phenotypic states highlights the strength of EUV ptychography in capturing subtle structural and biophysical variations that are often inaccessible to label-dependent techniques. Crucially, this level of quantitative reliability is enabled by our novel purity-based axial distance calibration[50], which minimizes reconstruction biases and ensures accurate retrieval of both amplitude and phase. The combined analysis of phase and amplitude signals measured at very short EUV wavelengths enables the extraction of quantitative descriptors (e.g., roughness, scattering density) that sensitively reflect nanoscale damage phenomena and form the basis for unsupervised clustering, allowing the identification of subclusters along a gradient of damage.

Such capability opens up new possibilities in microbiological research, including the study of antimicrobial mechanisms and host–microbe interactions[85] in-depth at unprecedented resolution. The informative approach could also contribute to future phenotypic screening pipelines, where quantitative structural readouts complement molecular or genetic assays for assessing treatment efficacy. Further, the multivariate datasets acquired by EUV ptychography enable analysis and discovery of potential associations and temporal relationships between different response phenotypes.

From a technical perspective, the extension of EUV ptychography toward shorter wavelengths, particularly into the water window region (wavelength 2.3 – 4.4 nm), offers compelling opportunities for native-state biological imaging[28,30,86]. The inherent contrast between water and carbon in this spectral range enables visualization of hydrated specimens without labeling or sectioning, bridging the gap between dry-state ultrastructure and physiologically relevant hydrated environments. Realizing such capabilities on a table-top platform, however, will require further developments in coherent source power[87-89], sample delivery and presevation[90], and dose-efficient imaging strategies. In particular, although no radiation-induced damage was observed at the approximately $10^5$ Gy dose applied in this work, further dose-efficient measures will be critical for maintaining the high-fidelity quantitative imaging of hydrated and radiation-sensitive biological systems.

The presented study demonstrates the unique potential of EUV ptychography to provide structural and biophysical insights into bacterial systems, offering a quantitative view into physiological and pathological transitions, and to complement conventional imaging techniques by its intrinsic element-specificity at nanometric resolution. As compact EUV and soft X-ray ptychographic microscopes continue to evolve, we anticipate a growing role for these techniques in microbiology and medical diagnostics where nanoscale structure and function intersect.

## Methods

### ***Bacterial culture conditions and sample preparation***

Bacteria were obtained from the German Collection of Microorganisms and Cellculture (DSMZ). *Escherichia coli* (DSM 423) and *Bacillus subtilis* (DSM 1090) were inoculated in Tryptic-Soy-Yeast Medium (TSY) at 37°C under shaking conditions. After a precultivation overnight, the bacteria were given to a fresh medium for 3 hours before preparation for measurements. To avoid artifacts in the measurements, the medium was washed off the surface of the bacteria by spinning down 1 ml of bacterial solution and resuspending the pellet in the same amount of distilled water. This washing step was performed three times. To increase the density of bacteria in the final solution, bacterial pellets were dissolved in 200 µl of distilled water.

To investigate the influence of monazomycin on *B. subtilis*, the sample cultivation was kept the same, with the only difference being the media used after the precultivation. Here, monazomycin was added to TSY with a final concentration of 10 µg/ml.

The sporulation of *B. subtilis* was achieved by the cultivation of the bacteria in Lysogeny Broth (LB) with the addition of 10 mg/l $MnSO_4 \cdot xH_2O$[62]. The cultivation took place in liquid LB at 37°C under shaking conditions for 11 days. Sample preparation was done in the same way as it was described above. After the sample preparation, 1 µl of the bacteria was dropped onto a SiN membrane and dried at room temperature. While classical sporulation protocols such as Difco medium can yield spores within 24 hours, we followed a simplified protocol for ease of implementation and sample preparation, rather than for optimizing sporulation efficiency.

### *High harmonic generation*

The high harmonic generation process is driven by a fiber-based chirped-pulse amplifier operating at a central wavelength of 1030 nm. The amplified pulses are compressed to < 7 fs by cascaded noble-gas-filled hollow-core fibers with a residual average power of 30 W at a repetition rate of 75 kHz and a pulse energy of 400 µJ. These few-cycle pulses are directed into a vacuum chamber and focused onto a 400 µm diameter gas jet with a backing pressure of 700 mbar argon. A broad EUV continuum is generated, with a photon flux of $7 \times 10^9$ photons/s/eV at 92 eV. The separation of the generated harmonics from the high-power driving laser is realized by four grazing incidence plates (GIPs). Afterward, the residual IR laser is entirely blocked utilizing two 100 nm zirconium (Zr) foils, which results in a transmission window between 70 eV and 120 eV. More details on the HHG source can be found in our earlier work[91].

### *EUV ptychographic microscope and workflow*

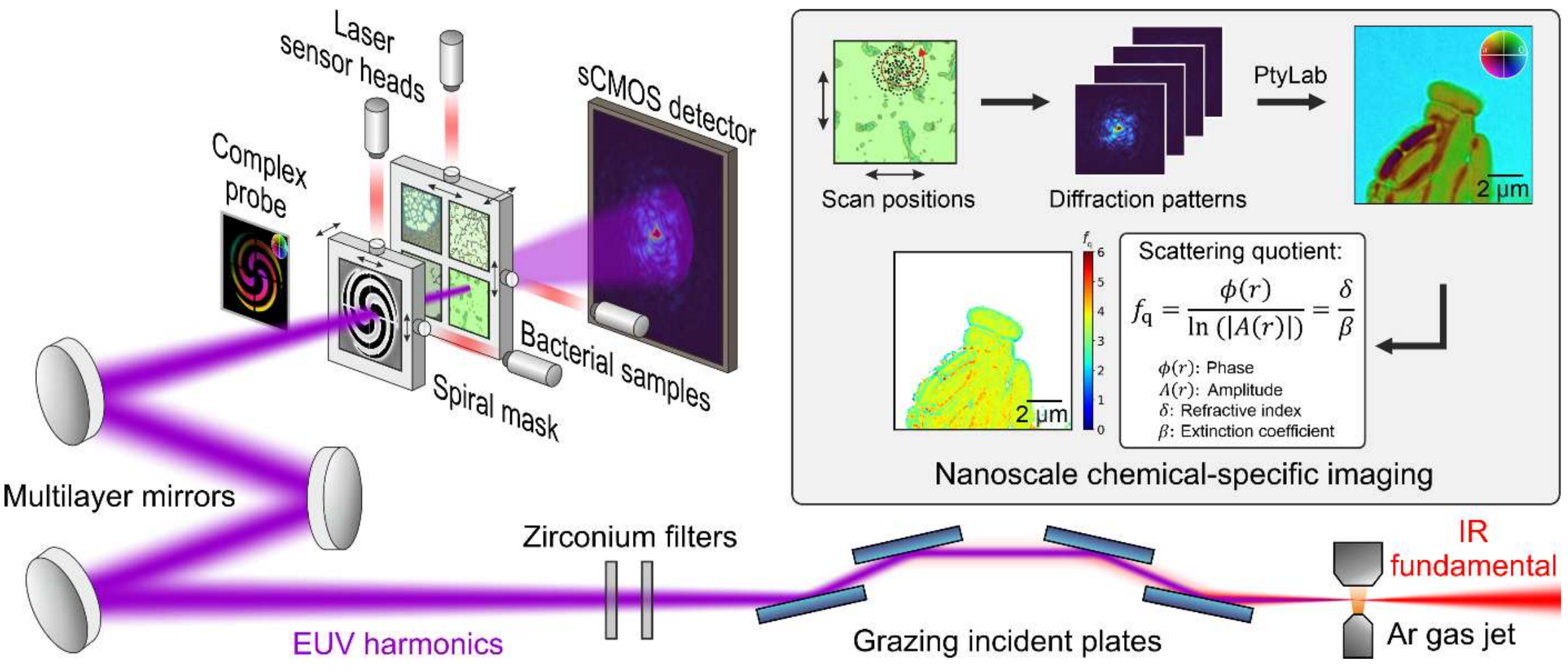

**Figure 6 Schematic of the EUV ptychographic microscope and imaging workflow.** A few-cycle IR laser is focused onto an argon gas jet, generating a broad EUV continuum. A narrow bandwidth of 0.2 nm is selected and focused at a wavelength of 13.5 nm using three multilayer mirrors. The structured light produced by a spiral mask illuminates the sample, and the resulting diffraction patterns are captured by a sCMOS detector during a ptychographic scan. The ptychographic iterative engine (PIE) enables amplitude and phase reconstruction. The calculation of the scattering quotient allows chemical-specific imaging.

The EUV ptychographic microscope has been previously described in a similar configuration[48]. Figure 6 illustrates the microscope setup and imaging workflow. The generated broadband EUV beam is spectrally filtered and focused onto a sample using reflective multi-layer mirrors. Different angles of incidence (AOI) on each mirror are used to compensate for astigmatism, yielding a combined peak reflectivity of ~23% and a spectral bandwidth (FWHM) of 0.2 nm at 13.5 nm. A binary spiral amplitude mask (transmission: 10% ~ 20%, depending on the mask position) is placed on a 3D positioner (SmarAct, Germany). Mask-based structural illumination has been demonstrated to enrich the spatial-frequency spectrum of the illumination, enhancing the signal-to-noise ratio (SNR) and spatial resolution of ptychographic reconstructions[92]. At the sample plane, the EUV photon flux of approximately $10^7$ photons/s with a relative spectral bandwidth of ~1.5 % at 92 eV. The imaged bacteria are adhered to a 50 nm thick SiN membrane (Norcada, Canada), positioned on a 3D stage a few hundred microns downstream from the masks. For precise positioning, laser interferometric stabilization (Picoscale, SmarAct GmbH, Germany) is applied to both the sample and mask stages, ensuring nanometer-scale accuracy during ptychographic scans. After the initial biological imaging demonstration, several technical upgrades were applied to the EUV ptychographic microscope. The integration of a high-speed sCMOS detector (GSENSE400BSI sensor, Andor Marana-X11, 20482048 pixels, pixel size 11 μm) and upgraded EUV focusing optics increased the photon flux by 50%, enabling faster acquisition times and larger fields of view[49]. The sCMOS detector is placed 32.7 mm downstream of the sample. Additionally, we implemented a fast and accurate purity-based self-calibration for axial distances within the momentum-accelerated PIE algorithm[93] in the open-source PtyLab[94] framework. This procedure improves axial distance accuracy and thereby enhances the reliability of quantitative, complex-valued reconstructions.

Chemical-specific imaging is achieved by calculating the scattering quotient ($f_q$), defined as the ratio of the real and imaginary components of the energy-dependent complex refractive index[28], as illustrated in the equation shown in Figure 6. The complex refractive index values required for the reference $f_q$ calculation at 92 eV can be obtained from the CXRO database[42]. For heterogeneous biological samples, the measured $f_q$ represents a composition-weighted effective ratio

$$\sum_k c_k f_{2,k} \Big/ \sum_k c_k f_{1,k}$$

where $c_k$ denotes the local concentration of constituent $k$, and $f_{1,2}$ the real and imaginary atomic scattering factor, respectively. Although not a single-material value, this effective quantity remains monotonically linked to biochemical composition and therefore provides robust relative contrast for mapping ultrastructural and chemical variation in biological specimens.

***Purity-based calibration on the sample-detector axial distance***

To minimize reconstruction bias and ensure accurate retrieval of high-fidelity images, we employ a purity-scan-based axial distance calibration procedure[50] for every measurement. This approach monitors a coherence-related metric, the probe purity μ, defined as[95]

$$\mu = \frac{\sqrt{\sum_n \lambda_n^2}}{\sum_n \lambda_n}$$

Where $\lambda_n$ denotes the singular values from the modal decomposition of the reconstructed probe. By scanning the axial distance and identifying the setting that maximizes μ, we achieve optimal forward-model matching and robust and optimal image retrieval. A validation example using the reconstruction of the *B. subtilis* spore is detailed in the Supplementary.

This calibration strategy eliminates reliance on ambiguous sharpness metrics or references, significantly enhancing the quantitative accuracy and reconstruction fidelity. It is especially effective for weak-scattering, soft biological specimens, and therefore represents an enabling capability for quantitative, subcellular imaging of bacterial morphology and composition.

***Data collection***

Multiple imaging datasets are obtained from independent bacterial samples, as illustrated in Supplementary Figure S2. Ptychographic scans are performed on various regions of the sample, using a 20 µm spiral grid size for systematic data collection. Each measurement employed high-dynamic-range (HDR) imaging and ensured reconstructions with sub-50 nm resolution and an imaging speed of approx. 2 Mpixel/h. The radiation dose per measurement was estimated to be approximately $10^5$ Gy. No dose-induced structural change was observed between repeated scans of an identical FOV, confirming that the applied dose did not introduce measurable damage under our experimental conditions. Detailed experimental parameters are provided in the Supplementary.

**Data Availability**

The data supporting the plots in this paper and other findings of this study are available in Ref. [96].

**Code Availability**

All algorithms used in this study are based on publicly available software libraries, including PtyLab for ptychographic reconstruction and standard Python packages for data processing. Custom scripts used for data visualization and statistical analysis are available in Ref. [96].

**Acknowledgments**

The research was sponsored by the Innovation Pool of the Research Field Matter of the Helmholtz Association of German Research Centers (project FISCOV), the Thüringer Ministerium für Bildung, Wissenschaft und Kultur (2018 FGR 0080), the Helmholtz Association (incubator project Ptychography 4.0), the Fraunhofer-Gesellschaft (Cluster of Excellence Advanced Photon Sources), the Free State of Thuringia (TAB; Multi-XUV/2023 FGR 0054), the German Funding Agency (Deutsche Forschungsgemeinschaft; under Germany´s Excellence Strategy – EXC 2051 –

Project-ID 390713860; Project number 316213987 – SFB 1278; GRK M-M-M: GRK 2723/1 – 2023 – ID 44711651), the Leibniz Association (Leibniz Collaborative Excellence Programme, project AMPel – project number K548/2023). Further, this work is supported by the BMFTR(Federal Ministry of Research, Technology and Space), funding program Photonics Research Germany (FKZ: 13N15713 / 13N15717) and is integrated into the Leibniz Center for Photonics in Infection Research (LPI). The LPI initiated by Leibniz-IPHT, Leibniz-HKI, UKJ and FSU Jena is part of the BMFTR national roadmap for research infrastructures.

**Author Contributions**

C.L. and L.L. performed the EUV ptychographic imaging experiments. C.L. analyzed the data. C.L., L.L., C.W., W.E., D.P.M., C.E., and J.R. interpreted the results. C.L. designed the probe mask and sample holder. C.W. and F.H. prepared the biological specimens: the bacteria (*E. coli, and B. subtilis*), monazomycin-treated *B. subtilis,* and *B. subtilis* sporulation sample. S.H.C. acquired the SEM images for validation. C.L., C.W., and S.H.C. wrote the manuscript. All authors discussed and contributed to the writing of the manuscript. J.R., C.E., and J.L. developed the initial idea and supervised the project.

**Competing interests**

The authors declare that they have no known competing financial interests or personal relationships that could have influenced the work reported in this paper.